# The wave-particle duality – does the concept of particle make sense in quantum mechanics? Should we ask the second quantization?

*Sofia D. Wechsler*

**Abstract**

The quantum object is in general considered as displaying both wave and particle nature. By particle is understood an item localized in a very small volume of the space, and which cannot be simultaneously in two disjoint regions of the space. By wave, to the contrary, is understood a distributed item, occupying in some cases two or more disjoint regions of the space.

The quantum formalism did not explain until today the so-called "collapse" of the wave-function, i.e. the shrinking of the wave-function to one small region of the space, when a macroscopic object is encountered. This seems to happen in "which-way" experiments. A very appealing explanation for this behavior is the idea of a particle, localized in some limited part of the wave-function.

The present article challenges the concept of particle. It proves in base of a variant of the Tan, Walls and Collett experiment, that this concept leads to a situation in which the particle has to be simultaneously in two places distant from one another – situation that contradicts the very definition of a particle. Another argument is based on a modified version of the Afshar experiment, showing that the concept of particle is problematic.

The concept of particle makes additional difficulties when the wave-function passes through fields. An unexpected possibility to solve these difficulties seems to arise from the cavity quantum electrodynamics studies done recently by S. Savasta and his collaborators. It involves virtual particles. One of these studies is briefly described here. Though, experimental results are expected, so that it is too soon to draw conclusions whether it speaks in favor, or against the concept of particle.

**Abbreviations**
1PWF  = single-particle wave-function
dBBI  = de Broglie-Bohm interpretation
EBS   = end-beam-splitter
GRW   = Ghirardi, Rimini, Weber
QM    = quantum mechanics
U.V.  = ultra violet

## 1. Introduction

The quantum object is in general considered as displaying both wave and particle nature. A particle is defined as an object localized in a very small volume of the space, so, it cannot be simultaneously in two places. To the contrary, a wave is a distributed object, occupying in some cases two or more disjoint regions of the space. These two types of behavior contradict one another. Though there is quite a consensus in the quantum community that the quantum object behaves in both ways, fits both pictures. Albert Einstein wrote:

> "It seems as though we must use sometimes the one theory and sometimes the other, while at times we may use either. . . . We have two contradictory pictures of reality; separately neither of them fully explains the phenomena of light, but together they do." [1]

Niels Bohr, in an account about a lecture he gave at the International Physical Congress at Como in September 1927, stressed the idea that one and the same quantum object displays particle properties in one experiment, but wave-properties in another experiment, depending on the experimental configuration, though the object has to be considered as having both features:

> "This crucial point . . . implies the impossibility of any sharp separation between the behaviour of atomic objects and the interaction with the measuring instruments which serve to define the conditions under which the phenomena appear. . . . Consequently, evidence obtained under different experimental conditions cannot be comprehended within a single picture, but must be regarded as complementary in the sense that only the totality of the phenomena exhausts the possible information about the objects." [2]

In direct relation with Bohr's explanations, J. A. Wheeler asked the daring question whether it could be that the quantum object feels in some way the locations of the apparatuses, beam-splitters, detectors, before actually touching them, and adopts, according to that, the behavior of a particle or of a wave. He proposed a series of experiments known under the name *delayed choice experiments* [3]. Many experiments were done attempting to answer Wheeler's question, e.g. [4 – 7] (see additional references in [8]), but with non-conclusive result. No clear evidence was found that the pre-measurement behavior was definitely of a particle, or of a wave. To the contrary, S. Afshar performed a series of experiments [9 – 11] from which he concluded that both types of behavior appear in each trial and trial of the experiment.

L. de Broglie [12, 13] and then D. Bohm [14] gave to the wave-particle dualism a mathematical formulation known in the literature as the de Broglie-Bohm interpretation (dBBI) of the quantum mechanics (QM). Their main idea was that the QM admits a sub-formalism, i.e. the quantum object consists in a particle moving on a continuous trajectory, with a velocity derived from the wave-function. However, despite the positive fact of removing the need of the enigmatic postulate of *collapse of the wave-function*, dBBI was proved wrong. The continuous trajectories of the particles were proved incompatible with the theory of relativity [15, 16]. Recently, it was proved that even without invoking the relativity such trajectories are incompatible with the experiment – section 5 in [17]. The section 3 of the present article presents a simplified version of the proof in [17]. L. de Broglie himself had doubts about how to define the concept of particle:

> "*We shall see that we must at all cost hold to the view that the intensity of the $\psi$-wave measures the probability of occurrence of the particle, even if our effort makes us sacrifice the traditional idea which gives to the particles a position, a velocity and a well-defined path.*" ( [13] page 87)

Independently of accepting or refuting the dBBI, the wave-particle duality continues to be taken for granted by most of the physicists still at present. Vis-à-vis this situation, Ghirardi, Rimini and Weber (GRW) proposed another interpretation of QM, which does not contain at all the concept of particle, but speaks of localization of the wave-function, by shrinking to a small volume in space [18].[1] The proposal is based on the known experimental fact that the particle behavior appears when the wave-function meets a macroscopic object. As clearly expressed by a follower of GRW,

> "Our experience in the use of quantum theory tells us that the state reduction postulate should not be applied to a microscopic system consisting of a few elementary particles until it interacts with a macroscopic object such as a measuring device." [20]

The GRW interpretation is not a complete theory, major improvements are desirable, but its basic idea, presented above, is correct. Moreover, it is in line with Feynman's explanation on the classical limit of QM:

---

[1] In fact, the original GRW article was [19], however, the present author has a strong criticism on that article, which speaks of quantum states of macroscopic objects. It is known from the example with Schrödinger's cat that a macroscopic object cannot be in a quantum superposition of states, so, the quantum description is unsuitable for such objects. For this reason that the reader is referred to the later article [18], instead. Although in [18] still appear sections which repeat the mistake in [19], but those are isolated sections that the reader may skip.

> "The classical approximation, however, corresponds to the case that the dimensions, masses, times, etc., are so large that *S* is enormous in relation to ℏ (= 1.05×10$^{-27}$erg·sec). Then the phase of the contribution *S*/ℏ is some very, very large angle. . . . small changes of path will, generally, make enormous changes in phase, and our cosine and sine will oscillate exceedingly rapidly between plus and minus values. The total contribution will then add to zero; . . . But for the special path $\bar{x}(t)$, for which *S* is an extremum, a small change in the path produces, in the first order at least, no change in *S*. All the contributions from the paths in this region are nearly in phase, . . . , and do not cancel out"[2] [21]

Of course, it is not sufficient to propose an interpretation of the QM based only on waves, without explaining why the concept of particle is problematic. Different authors expressed doubts about this concept, as for instance C. Blood who formulated a general message to physicists to check if this concept is really needed:

> "it seems awkward to have a two-tiered scheme in which wavefunction-based quantum mechanics determines all the numbers, while particles—absent from the quantum mathematics—supply the structure necessary for agreement with our perceptions. This suggests we take a close look to see if particles are really needed." [22]

(More references can be found in [23]). Some physicists made attempts to rule out this concept by rigorous proofs. The theorems of Hegerfeldt [24, 25] and Malament [26], meant to prove that the concept of particle is at odds with the theory of relativity, triggered a whole debate (see a discussion in [27]). The present author also has doubts about the proofs in [24, 25, 26].[3]

The present article presents a proof against the concept of particle, in base of the analysis of a variant of the Tan, Walls and Collett (TWC) experiment [28]. The proof is simple, to the difference from those in [24, 25, 26]. It is proved that if the concept of particle is correct, then, in this experiment the particle should be present simultaneously in two places, which contradicts the very definition of this concept, given in the beginning of this section. The non-plausibility of this concept is also exemplified on a modified version of Afshar's experiments [9 – 11].

An additional problem with this concept appears when a single-particle wave-function (1PWF), possessing more than one wave-packet, passes through fields. For instance, assume that the wave-function describes an electron, has two wave-packets, and each wave-packet passes through an electric field. The result is that both wave-packets are accelerated. Then, each wave-packet carries an electron? That is impossible, this is a single-particle wave-function.

Recently, Garziano et al. [29] published a theoretical study in which appears that a quantum system may undergo a process where, between the initial and the final state, virtual particles may intervene violating the energy conservation. That would offer an interesting answer to the above question. But, on the other hand, according to [29] the process of absorption of a quantum system by an atom is more complicated than the non-relativistic QM describes. By the time this article is written, and as far as it is known to the present author, no experimental confirmation of the calculi in [29] is available. Therefore, it is too soon to draw a firm conclusion whether [29] speaks in favor, or against the concept of particle.

In continuation, the section 2 describes a which-way experiment illustrating the fact that such experiments invite the idea of particle. Section 3 examines a variant of the TWC experiment and proves that this idea leads to a contradiction. Section 4 presents a modified version of the Afshar experiment [10], and reveals a problem

---

[2] *S* is the action function.

[3] Both [24] and [25] contain the proof of a theorem which concludes that the concept of particle comes to a contradiction with the theory of relativity. The final part of the theorem questions the analyticity of the Fourier transform of the product of two functions, and shows that a contradiction appears. But the two functions have disjoint supports, therefore their product is null at any point in space. Thus, it is not clear how could it be that the Fourier transforms of a null function can have analyticity problems.

Malament's proof uses a lemma – page 7 in [26] – based on two assumptions, (i) and (ii). Checking the assumption (ii), the present author found that it may be valid only at one single time, not over a whole interval of time, as Malament assumed.

with the concept of particle. Section 5 points to an additional problem with this concept and brings into discussion new predictions of the second quantization. Section 6 contains conclusions.

## 2. Are which-way experiments a testimony that the quantum object has particle-nature?

Consider a 1PWF, e.g. the signal photon from a down-conversion pair, passing through a 50%-50% beam-splitter, and then a series of movable mirrors arrange the two resulting wave-packets $|1;a\rangle$ and $|1;b\rangle$ in line, one after the other, figure 1. The idler photon is sent to a separate detector I. Let the distances in the apparatus be so that by the time the detector I reports a recording, the wave-packet $|1;a\rangle$ reaches the detector S.

In a trial of the experiment in which $|1;a\rangle$ is recorded, it is known for sure that $|1;b\rangle$ won't produce a detection, and vice-versa. The wave-function of the signal photon is, according to the second quantization:

$$|\psi\rangle = (-1/\sqrt{2})(|1;a\rangle|0;b\rangle + i|0;a\rangle|1;b\rangle). \tag{1}$$

This wave-function reads as follows: if the detector S reports the wave-packet $|1;a\rangle$, only *void* impinges in continuation on S; however, equation (1) shows that there exists an alternative wave-packet $|1;b\rangle$, *non-void*, which also impinges on S. Though, it won't trigger S if $|1;a\rangle$ did. Alternatively, if $|1;a\rangle$ doesn't trigger a recording in S, $|1;b\rangle$ would do it for sure, although $|1;a\rangle$ is the first wave-packet to meet S.

This is why the idea of a *particle* contained in one of the wave-packets is so appealing. The other wave-packet is then considered "empty wave", i.e. it carries no particle; though it possesses all the properties of the respective type of particle, as charge, polarization, etc., so that it can participate in interference.

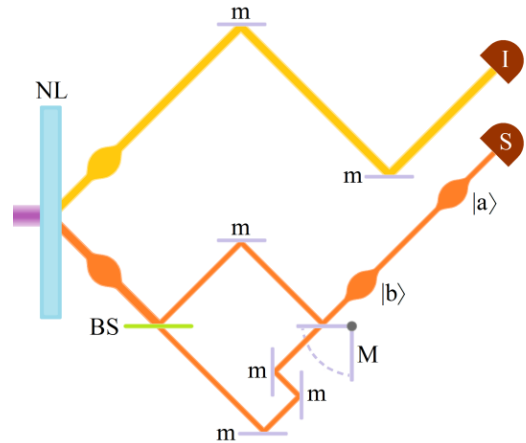

Figure 1. Two wave-packets from a 1PWF, arranged in line.
NL is a nonlinear crystal splitting ultra-violet photons into signal-idler pairs. The idler photon is sent to a detector I the click of which heralds the presence of the signal photon in the apparatus. The wave-packet of the latter is split by the 50%-50% beam-splitter BS, into a reflected and a transmitted copy, $|1;a\rangle$, respectively $|1;b\rangle$. m and M are mirrors, m are fixed and M is rotatable. When $|1;a\rangle$ reaches the mirror M, this mirror is in horizontal position, redirecting $|1;a\rangle$ towards the detector S. Then, M is rotated, so that the wave-packet $|1;b\rangle$, which comes later, continues its travel towards S.

The idea of empty waves raises question. A wave-packet of frequency $\nu$ carries, by Planck's formula, an energy $h\nu$, therefore there is no obvious reason why this energy should be not delivered to a detector.
The question whether empty waves can exist preoccupied the physicists and fueled a whole debate toward the end of the previous century – [30 – 41] are a couple of proposals for experiments and exchanges of opinions.

## 3. A variant of the Tan-Walls-Collett experiment – where is the particle?

Consider a 1PWF from a source S. The wave-packet passes through a beam-splitter BS, which reflects 1/3 from the incident intensity. The transmitted wave-packet passes through a second beam-splitter, BS', which reflects and transmits in equal proportion – figure 2. In continuation, each wave-packet travels to a 50-50% end-beam-splitter (EBS), BS$_1$, BS$_2$, and BS$_3$, respectively. On the path of each beam is placed a phase-shifter, $\theta_1$ on the path **a**, $\theta_2$ on the path **b**, and $\theta_3$ on **c**. Each path from BS to the EBS is of equal length, only the phase-shifters introduce differences in the wave-packets phases. The resulting wave-function is

$$|\psi\rangle = (-1/\sqrt{3})\left(e^{i\theta_1}|1;\mathbf{a}\rangle|0;\mathbf{b}\rangle|0;\mathbf{c}\rangle + |0;\mathbf{a}\rangle e^{i\theta_2}|1;\mathbf{b}\rangle|0;\mathbf{c}\rangle + |0;\mathbf{a}\rangle|0;\mathbf{b}\rangle e^{i\theta_3}|1;\mathbf{c}\rangle\right). \qquad (2)$$

In this expression the phase accumulated by each one of the wave-packets $|1;\mathbf{a}\rangle$, $|1;\mathbf{b}\rangle$ and $|1;\mathbf{c}\rangle$, from BS to the EBS was omitted as being the same for all three wave-packets. Only the additional phase-shifts $\theta_1$, $\theta_2$, and $\theta_3$, respectively, were written explicitly.

On the other side of each end-beam-splitter impinges a coherent beam of the form

$$|\alpha_j\rangle = \mathcal{N}\left(|0\rangle + q|1;\mathbf{e}_j\rangle + \ldots\right), \qquad j = 1, 2, 3. \qquad (3)$$

where $\mathcal{N}$ is the normalization factor and $q$ is a complex number with $|q| < 1$, both $\mathcal{N}$ and $q$ being the same for all three coherent beams. All the photons are of the same polarization, so, we don't write it explicitly. The total wave-function at this step is

$$|\Psi\rangle = (-\mathcal{N}^3/\sqrt{3})\left(e^{i\theta_1}|1;\mathbf{a}\rangle|0;\mathbf{b}\rangle|0;\mathbf{c}\rangle + |0;\mathbf{a}\rangle e^{i\theta_2}|1;\mathbf{b}\rangle|0;\mathbf{c}\rangle + |0;\mathbf{a}\rangle|0;\mathbf{b}\rangle e^{i\theta_3}|1;\mathbf{c}\rangle\right)\prod_{j=1}^{3}\left(|0\rangle + q|1;\mathbf{e}_j\rangle + \ldots\right). \qquad (4)$$

*Convention*: since the expressions written according to the 2$^{nd}$ quantization are very long, we will omit in the rest of this section the void wave-packets in the formulas.

We will be interested here only in the trials ending with a triple detection, one detection at the output of each EBS. Opening the parentheses in (4) and writing explicitly only the suitable terms, one gets [4]

$$|\Psi\rangle = -\mathcal{M}\left\{\left(e^{i\theta_1}|1;\mathbf{a}\rangle|1;\mathbf{e}_2\rangle|1;\mathbf{e}_3\rangle + e^{i\theta_2}|1;\mathbf{e}_1\rangle|1;\mathbf{b}\rangle|1;\mathbf{e}_3\rangle + e^{i\theta_3}|1;\mathbf{e}_1\rangle|1;\mathbf{e}_2\rangle|1;\mathbf{c}\rangle\right) + \ldots\right\}, \quad \mathcal{M} = \mathcal{N}^3 q^2/\sqrt{3}. \qquad (5)$$

---

[4] Cases in which beyond each EBS clicks one detector, however on a detector land two or more photons, are of small probability, since $|q| < 1$ and their amplitudes of probability are proportional with $q^n$ where $n > 3$. Also, the cases of triple detection caused by $|1;\mathbf{e}_1\rangle|1;\mathbf{e}_2\rangle|1;\mathbf{e}3\rangle$ have the amplitude of probability proportional with $q^3$ instead of $q^2$, so, are much less probable.

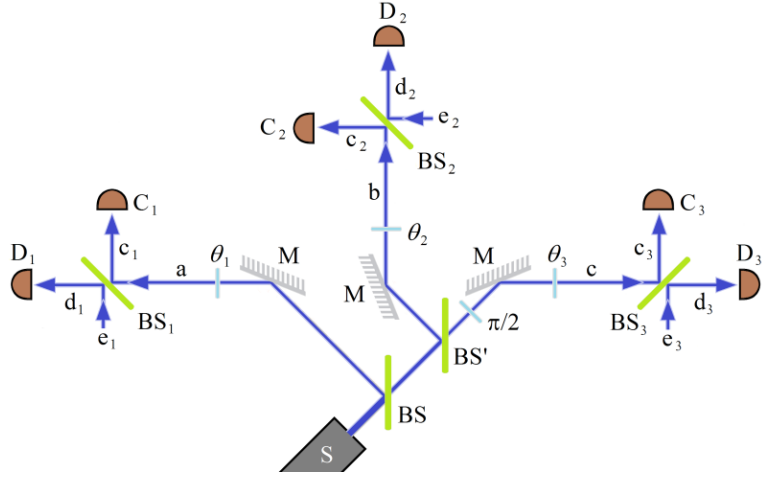

Figure 2. A single-particle wave-function showing a simultaneous effect in three places.
See explanations in the text.

At the EBSs take place the transformations

$$|1;\mathbf{u}_j\rangle \to (1/\sqrt{2})\left(i|1;\mathbf{c}_j\rangle + |1;\mathbf{d}_j\rangle\right), \quad |1;\mathbf{e}_j\rangle \to (1/\sqrt{2})\left(|1;\mathbf{c}_j\rangle + i|1;\mathbf{d}_j\rangle\right), \quad j = 1, 2, 3, \quad (6)$$

where $\mathbf{u}_1 = \mathbf{a}$, $\mathbf{u}_2 = \mathbf{b}$, and $\mathbf{u}_3 = \mathbf{c}$. Introducing (6) in (5) there results after some long, though simple calculus

$$|\Theta\rangle = \mathcal{M}\left\{\frac{1}{\sqrt{8}}\left[\left(e^{i\theta_1} + e^{i\theta_2} + e^{i\theta_3}\right)\left(-i|1;\mathbf{c}_1\rangle|1;\mathbf{c}_2\rangle|1;\mathbf{c}_3\rangle + |1;\mathbf{d}_1\rangle|1;\mathbf{d}_2\rangle|1;\mathbf{d}_3\rangle\right)\right.\right.$$
$$\left.\left.+ \sum_{j,k,l \in \{1,2,3\}}\left(e^{i\theta_j} + e^{i\theta_k} - e^{i\theta_l}\right)\left(|1;\mathbf{c}_j\rangle|1;\mathbf{c}_k\rangle|1;\mathbf{d}_l\rangle - i|1;\mathbf{d}_j\rangle|1;\mathbf{d}_k\rangle|1;\mathbf{c}_l\rangle\right)\right] + \ldots\right\}. \quad (7)$$

The joint probabilities of clicks in three detectors are

$$\text{Prob}(C_1 \& C_2 \& C_3) = \text{Prob}(D_1 \& D_2 \& D_3) = \frac{|\mathcal{M}|^2}{8}\left|e^{i\theta_1} + e^{i\theta_2} + e^{i\theta_3}\right|^2$$
$$\text{Prob}(C_j \& C_k \& D_l) = \text{Prob}(D_j \& D_k \& C_l) = \frac{|\mathcal{M}|^2}{8}\left|e^{i\theta_j} + e^{i\theta_k} - e^{i\theta_l}\right|^2, \quad j,k,l \in \{1,2,3\}, \; j \neq k \neq l \neq j. \quad (8)$$

This result makes obvious the wave-nature of the photon, at least before detection. One can see that in each trial resulting in a triple detection as in (7), the phase-shifts $\theta_1$, $\theta_2$, and $\theta_3$ were present together, therefore so were $|1;\mathbf{a}\rangle$, $|1;\mathbf{b}\rangle$, and $|1;\mathbf{c}\rangle$, which carried the phase-shifts.

However, with the concept of *particle* there appear problems.
For showing this, we assume in the rest of this section that the concept is correct, and that the particle travels with some wave-packet. Thus, the particle exiting the source S – named below "particle S" – is assumed to travel with $|1;\mathbf{a}\rangle$, or $|1;\mathbf{b}\rangle$, or $|1;\mathbf{c}\rangle$, cross the respective EBS, and end up in one of the detectors beyond the EBS.

We will set the values of the phase-shifts to $\theta_1 = \theta_3 = 0$, and $\theta_2 = \pi$, and focus on the trials ending with a joint detection in the detectors $D_1$, $D_2$, and $D_3$. Calculating the amplitude of probability $\mathcal{A}(D_1 \& D_2 \& D_3)$, we will show that a contradiction appears.

For shortening the text, a product of three wave-packets will be named 3-wave.

Introducing in (7) the above values for the phase-shifts one gets

$$\mathcal{A}(D_1 \& D_2 \& D_3) = \mathcal{M}/\sqrt{8}. \tag{9}$$

Let's notice that in the RHS of (5) appear the 3-waves $|1;\mathbf{a}\rangle|1;\mathbf{e}_2\rangle|1;\mathbf{e}_3\rangle$, $|1;\mathbf{e}_1\rangle|1;\mathbf{b}\rangle|1;\mathbf{e}_3\rangle$, and $|1;\mathbf{e}_1\rangle|1;\mathbf{e}_2\rangle|1;\mathbf{c}\rangle$. Introducing in (5) the values of the phase-shifts, one finds that the amplitudes of probability of these 3-waves are equal to $-\mathcal{M}$, $\mathcal{M}$, and $-\mathcal{M}$, respectively. Therefore, due to the transformations (6) at the EBSs, their contributions to $\mathcal{A}(D_1 \& D_2 \& D_3)$ are as follows:

$$\mathcal{A}(D_1 \& D_2 \& D_3 \,|\, |1;\mathbf{a}\rangle|1;\mathbf{e}_2\rangle|1;\mathbf{e}_3\rangle) = \mathcal{M}/\sqrt{8}, \quad \mathcal{A}(D_1 \& D_2 \& D_3 \,|\, |1;\mathbf{e}_1\rangle|1;\mathbf{b}\rangle|1;\mathbf{e}_3\rangle) = -\mathcal{M}/\sqrt{8},$$
$$\mathcal{A}(D_1 \& D_2 \& D_3 \,|\, |1;\mathbf{e}_1\rangle|1;\mathbf{e}_2\rangle|1;\mathbf{c}\rangle) = \mathcal{M}/\sqrt{8}. \tag{10}$$

Comparing (10) with (9) one can see that the result in (9) is a consequence of the fact that two of the contributions of the 3-waves in (5) to $\mathcal{A}(D_1 \& D_2 \& D_3)$, cancel out mutually: either the contribution of $|1;\mathbf{a}\rangle|1;\mathbf{e}_2\rangle|1;\mathbf{e}_3\rangle$ and that of $|1;\mathbf{e}_1\rangle|1;\mathbf{b}\rangle|1;\mathbf{e}_3\rangle$, or the contribution of $|1;\mathbf{e}_1\rangle|1;\mathbf{e}_2\rangle|1;\mathbf{c}\rangle$ and that of $|1;\mathbf{e}_1\rangle|1;\mathbf{b}\rangle|1;\mathbf{e}_3\rangle$. From these 3-waves one can see also that if $|1;\mathbf{a}\rangle$ carries the particle S, then the detections in $D_1$, $D_2$ and $D_3$ are due to the particles carried by the wave-packets $|1;\mathbf{a}\rangle$, $|1;\mathbf{e}_2\rangle$, and $|1;\mathbf{e}_3\rangle$. If the wave-packet $|1;\mathbf{b}\rangle$ is the one that carries the particle S, then the detections in $D_1$, $D_2$ and $D_3$ are caused by particles carried by $|1;\mathbf{e}_1\rangle$, $|1;\mathbf{b}\rangle$, and $|1;\mathbf{e}_3\rangle$. If $|1;\mathbf{c}\rangle$ carries the particle S, the particles are carried by $|1;\mathbf{e}_1\rangle$, $|1;\mathbf{e}_2\rangle$, and $|1;\mathbf{c}\rangle$.

These conclusions can be summarized as follows:
1. If the three particles are in $|1;\mathbf{a}\rangle$, $|1;\mathbf{e}_2\rangle$, and $|1;\mathbf{e}_3\rangle$, respectively, the entire amplitude $\mathcal{A}(D_1 \& D_2 \& D_3)$ is contributed by the 3-wave $|1;\mathbf{a}\rangle|1;\mathbf{e}_2\rangle|1;\mathbf{e}_3\rangle$ – compare the first equality in (10) with (9).
2. If the three particles are in $|1;\mathbf{e}_1\rangle$, $|1;\mathbf{e}_2\rangle$, and $|1;\mathbf{c}\rangle$, respectively, the entire amplitude $\mathcal{A}(D_1 \& D_2 \& D_3)$ is contributed by the 3-wave $|1;\mathbf{e}_1\rangle|1;\mathbf{e}_2\rangle|1;\mathbf{c}\rangle$ – compare the last equality in (10) with (9).

From **1** it immediately results that for obtaining a joint detection in $D_1$, $D_2$ and $D_3$, the particle S should travel with the wave-packet $|1;\mathbf{a}\rangle$ and end up in the detector $D_1$. However, from **2** it results that for obtaining this joint detection the particle S should travel with the wave-packet $|1;\mathbf{c}\rangle$ and end up in $D_3$.

We got an impossibility, the particle S can't end up at once in $D_1$ and $D_3$. A particle is a localized object.

One may suggest that in part of the trials ending with the joint detection in $D_1$, $D_2$, and $D_3$, the particle S took the path **a**, and in the other trials, the path **c**. But, this is not a solution, because each one of the 3-waves $|1;\mathbf{a}\rangle|1;\mathbf{e}_2\rangle|1;\mathbf{e}_3\rangle$ and $|1;\mathbf{e}_1\rangle|1;\mathbf{e}_2\rangle|1;\mathbf{c}\rangle$ contributes the *entire* amplitude $\mathcal{A}(D_1 \& D_2 \& D_3)$.

As a side remark, proving that a particle should be present in some cases in two places at once, rules out the Bohmian mechanics, which defines a the particle concept as in the present article, i.e. as a localized object.

## 4. Which relationship may exist between the supposed *particle* and the *wave*?

This section proposes and analyzes a modified version of Afshar's experiment in [10]. A diagonally-polarized (D) photon is sent upon a screen with two slits, labeled V and H. In front of the slit H (V) is placed half of a polarizing sheet which lets pass only photons polarized H (V). A convergent lens creates an image of the two slits on the openings of a double collimator C – figure 3a. A photographic plate S records the beams 1 and 2 exiting the collimator.

In the region of the lens the beams from the two slits overlap; in the vicinity of the symmetry plane $y = 0$, the wave-function of the photon can be approximated by

$$|\psi\rangle = \frac{e^{i\kappa_z z}}{\sqrt{2}}\left(e^{i\kappa_y y}|H\rangle + e^{-i\kappa_y y}|V\rangle\right) = e^{i\kappa_z z}\left[\cos(\kappa_y y)|D\rangle + i\sin(\kappa_y y)|A\rangle\right], \tag{11}$$

where $\kappa_z$ and $\kappa_y$ are the horizontal, respectively vertical, wave-number, and

$$|H\rangle = \frac{1}{\sqrt{2}}(|D\rangle + |A\rangle), \quad |V\rangle = \frac{1}{\sqrt{2}}(|D\rangle - |A\rangle). \tag{12}$$

By the rightmost wing of (11), on the planes $y = n\pi/\kappa_y$ the probability of finding the photon polarized A is vanishingly small, however the probability to find it polarized D is maximal.

On the plane $n = 0$, i.e. $y = 0$, is centered an opaque wire – figure 3b – with a tiny diameter of $\pi/(10\kappa_y)$.

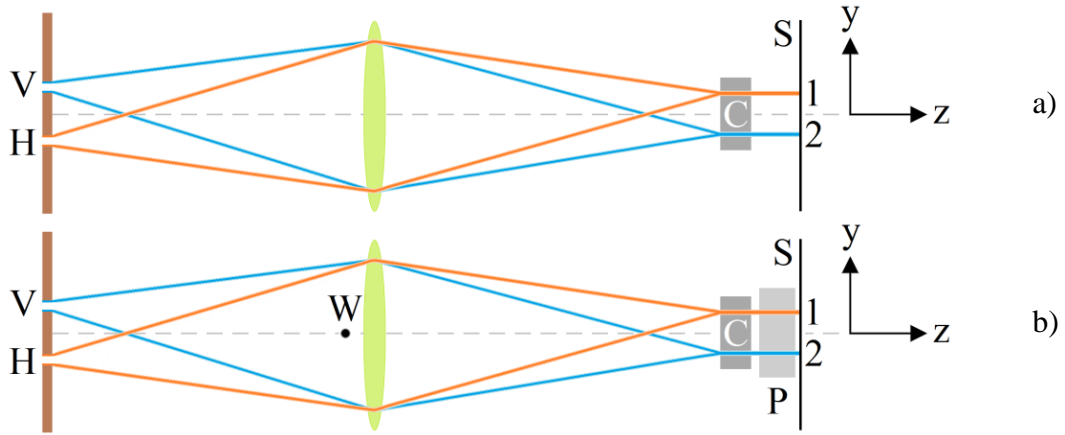

Figure 3. The Afshar experiment modified.
The experiment is described in the 2D geometry. The colors are only for eye-guiding – the wavelengths of the beams exiting the slits H and V are the same. The blue and orange colored rays, delimit (approximately) the beams exiting the slits. The dot W represents the wire, the light-green form represents the lens, C is the double collimator, P the polarizer, and S the photographic plate. 1 (2) is a region on the photographic plate where appears the image of the slit H (V).

The wire effect is to remove photons from the beam, leaving the wave-function in the form

$$|\Phi\rangle = \frac{e^{i\kappa_z z}}{\sqrt{2}} e^{-\gamma(y,z)} \left( e^{i\kappa_y y}|H\rangle + e^{-i\kappa_y y}|V\rangle \right) = e^{i\kappa_z z - \gamma(y,z)} \left[ \cos(\kappa_y y)|D\rangle + i\sin(\kappa_y y)|A\rangle \right], \quad (13)$$

where $\gamma(y,z)$ is a real and positive function which decreases rapidly to zero outside the volume of the wire. Therefore, close to the lens the intensity of illumination behaves as follows

$$\begin{aligned} \mathcal{J}(H; y, z) &= \tfrac{1}{2} e^{-2\gamma(y,z)}, & \mathcal{J}(V; y, z) &= \tfrac{1}{2} e^{-2\gamma(y,z)}, \\ \mathcal{J}(D; y, z) &= e^{-2\gamma(y,z)} \cos^2(\kappa_y y), & \mathcal{J}(A; y, z) &= e^{-2\gamma(y,z)} \sin^2(\kappa_y y). \end{aligned} \quad (14)$$

From the figure 3 one can see that in the absence of the polarizer P, the images 1 and 2 on the photographic plate are polarized H and V, respectively. Due to the polarizer the following effects are expected to occur:

- **i)** If P transmits only the polarization H (V), only the image 1 (2) would be seen on the photographic plate S. The factor $e^{-2\gamma(y,z)}$ entails an intensity of each one of these images weaker than in the absence of the wire.

- **ii)** If the sheet transmits only the polarization A, the two images 1 and 2 would be seen on S, because each one of the polarizations H and V contains a component A, as show the relations (12). It has to be mentioned that the wire would have a very small effect on the intensities of these images because the function $\gamma(y, z)$ is null in the region where the sine square is maximal – see (14).

- **iii)** If P transmits only the polarization D, both images 1 and 2 would be seen on S, because both polarizations H and V contain a component D. However, because of the wire, the intensities of the two images would be smaller than in the case **ii** because the function $\gamma(y, z)$ is maximal in the region $y = 0$ where the square cosine is maximal – see (14).

In relation with the wave-particle duality, the case **i** fits the idea of a *particle*. One can say that in a given trial of the experiment the particle comes either from the slit H, or from the slit V, and is polarized accordingly. So, if not absorbed by the wire, the particle reaches the polarizer P, and passes on if P transmits the same polarization as that of the photon.

However, **ii** and **iii** hint of a tableau of *waves*. The formulas of $\mathcal{J}(D; y, z)$ and $\mathcal{J}(A; y, z)$ in (14) indicate interference, and interference occurs between two waves, the waves coming from the slits. There is no hint of a particle behavior *before the detection on the plate S*.
But these two waves that interfere carry polarizations. Polarization is a property of an electrical field, and fields possess energy. Since we have to do with a 1PWF it may be assumed that the wave from one of the slits also carries the particle. Though, the energy, $\hbar\omega$, and the polarization, belong to the waves. Then it's not clear which physical properties does this particle possess. In base of which physical properties can it impress the detector?

The picture of fields possessing energy, which though is not delivered to detectors, or not felt by the detectors, and a particle about which it is not clear what physical properties possesses, but yes impresses a detector, *is a non-plausible picture.*

## 5. What tells us the second quantization?

There is an additional problem with the concept of particle. Consider a 1PWF representing an electron and possessing, say, two wave-packets. Assume that both wave-packets pass through electric fields, therefore, both are accelerated. That means that each wave-packet possesses the electron charge. But this is a problem because we have to do with a single-electron wave-function.

An experiment performed by Garziano et al. [29] with photons (not with electrons), showed a new effect of cavity quantum electrodynamics: one and the same photon can excite two atoms – actually, artificial atoms – situated at some distance from one another. The process passes through intermediate states – see figure 4 which reproduces two of the channels of the process. In the intermediary states between the initial state $|g,g,1\rangle$ and the final state $|e,e,0\rangle$, appear virtual photons violating the system energy conservation – the system consisting in the atoms and the photons in the cavity. States violating the energy conservation can't be detected and are named "virtual states". The system energy is conserved only between the initial and the final state.

Though, the virtual photons perform a practical task which can be observed indirectly: one can see in the figure 4 that at least one of the artificial atoms is excited by a virtual photon, not by a real one. The figure 3 in [29] shows that the system oscillates between the initial and the final state with a frequency $\Omega_{\text{eff}}$.

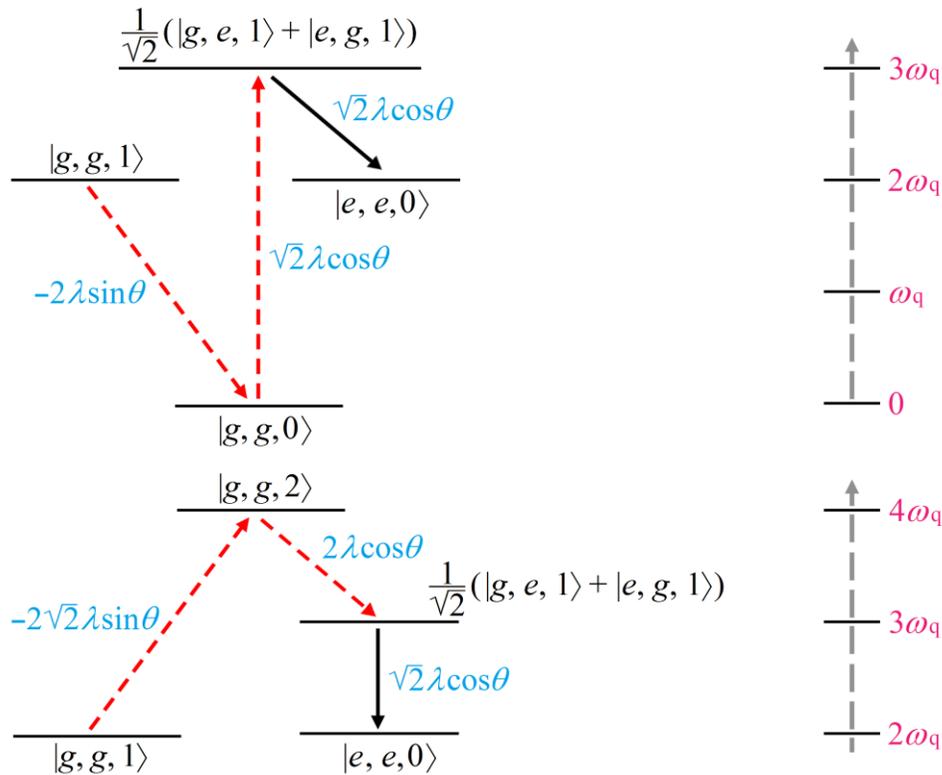

Figure 4. Two of the four channels of excitation of two atoms with one photon. '$g$' ('$e$') denotes the atom ground (excited) state, $\hbar\omega_q$ is the excitation energy of the atom (½ the photon energy). Red arrows lead to virtual states, and black arrows to the final state. The light blue expressions by the side of the arrows indicate the transition amplitudes between states. The scale on the right helps to see the total energy in each state.

This frequency is determined, according to the perturbation theory, by the transition amplitudes between the states through which the process passes, and by the energies of these states – equations (3 – 7) in [29],

$$\Omega_{\text{eff}} = -\sum_{m,n} \frac{V_{fn} V_{nm} V_{mi}}{(E_i - E_m)(E_i - E_n)}. \tag{15}$$

In this equality the summation is over all the channels of the process; the indexes *i* and *f* are for the initial and final state, *m* and *n* for the intermediate states which are *virtual* as said above, $E_j$ is the energy of the system (atoms + photons) in the state *j*, and $V_{kj}$ the transition amplitude from the state *j* to the state *k*.
Thus, if the practical experiment would confirm the formula (15), it would also confirm the existence of the virtual states.

It is clear that the real photon and the atoms do not "live" in absolute void, but in a restless sea of photons which pop up from the sea for an extremely short time and return to the sea. The sea may even absorb real photons and release them back, participating thus actively in quantum processes. It is very plausible that the sea contains additional types of quantum objects.

However, there is another surprising thing in this study, for which no physical explanation seems possible. In the 3$^{rd}$ state of each channel in figure 4, one of the atoms is excited with half of the photon energy. Where from does the atom take this energy? The cavity length is equal to $\lambda/2$, where $\lambda$ is the wavelength of the real photon, the same as that of the virtual photon. For containing a virtual photon of half energy, i.e. of double wavelength, the cavity should be twice longer.

Returning to our topic, the dualism, does this study speak in favor of the particle concept, or in disfavor? By the time the present article is written, it is not known to the author whether any results of a practical implementation of [29], was published. Thus, it is too soon to answer questions.

## 6. Conclusions

The present text examined the wave-particle duality and brought arguments against the concept of particle.
Section 1 presented a brief history of the attitude of the physicists toward this duality.
In the section 2 was examined an experiment which illustrates why the concept of particle is attractive. However, section 1 also stressed that the particle concept is relevant only at the encounter with a macroscopic object. Before this encounter, the quantum object behaves as a wave, fact especially illustrated by the sections 3 and 4.
The section 3 presented a rigorous proof that the concept of particle leads to a contradiction. The novelty of this proof stands in not using moving frames of coordinates, therefore not having to confront the challenge of the question whether a preferred frame exists – see a discussion of this problem in [42].
In the experiment examined in section 4, the case **i** may be satisfactorily explained by assuming the existence of a particle, as in the experiment in section 2. However, the cases **ii** and **iii** show that the case **i** does not tell us all the truth. The cases **ii** and **iii** reveal the existence of waves, which carry energy. On the other hand, it is not clear which physical properties are carried by the particle, in base of which the particle impresses a detector. The picture of waves possessing physical properties though not impressing a detector, and a particle with no clear physical property, though impressing a detector, is non-plausible. As exemplified in section 1, for explaining the process of localization of the wave-function there are additional proposals.

Section 5 points to one more problem. How many particles carries a 1PWF? It is argued that each wave-packet of the wave-function should carry a particle. That is impossible for a 1PWF.

A novel study in the domain of cavity quantum electrodynamics seems to suggest a solution to this problem, based on the intervention of virtual quantum systems. The study also puts in evidence a strange problem about the content of the quantum vacuum. But, whether this study speaks in favor of the idea of a particle, or in disfavor, it is too soon to say, as for the moment, results from practical implementation of the study are not yet known.

## Acknowledgements

I am grateful to Peter N. Tanguay for bringing to my attention the Afshar experiment, and to Ing. Stefano Quattrini for asking a question that made me examine the reference [29] and related material, in consequence of which I wrote the section 5 of the present article. I am also in great debt to the reviewer at the journal, for the very attentive examination of my article, and for the greatly helpful advices.